\documentclass[twocolumn,showpacs,preprintnumbers,amsmath,amssymb,aps,letterpaper,superscriptaddress]{revtex4}
\usepackage{graphicx}% Include figure files
\usepackage{dcolumn}% Align table columns on decimal point
\usepackage{bm}% bold math

\begin{document}

\preprint{APS/123-QED}

\title{Direct Evidence for Edge-Contaminated Vortex Phase in a Nb Single Crystal using Neutron Diffraction}% Force line breaks with \\
\author{H.A. Hanson}
 \affiliation{Department of Physics, Brown University, Providence RI 02912, USA}
\author{X. Wang}
    \affiliation{Department of Physics, Brown University, Providence RI 02912, USA}
\author{I.K. Dimitrov$^\ast$}
 \affiliation{Department of Physics, Brown University, Providence RI 02912, USA}
\author{J. Shi$^\dagger$}
    \affiliation{Department of Physics, Brown University, Providence RI 02912, USA}
\author{X.S. Ling}
    \affiliation{Department of Physics, Brown University, Providence RI 02912, USA}
 \email{xsling@brown.edu}
\author{B.B. Maranville}
 \affiliation{ NIST Center for Neutron Research, Gaithersburg, MD 20899, USA}
\author{C.F. Majkrzak}
 \affiliation{ NIST Center for Neutron Research, Gaithersburg, MD 20899, USA}
\author{M. Laver}
 \affiliation{Paul Scherrer Institut, Villigen PSI, CH 5232 Switzerland}
\author{U. Keiderling}
 \affiliation{Helmholtz Zentrum Berlin fr Materialien und Energie GmbH, 14109 Berlin,Germany}
\author{M. Russina}
 \affiliation{Helmholtz Zentrum Berlin fr Materialien und Energie GmbH, 14109 Berlin,Germany}
\date{\today}% It is always \today, today,
       % but any date may be explicitly specified
\begin{abstract}
\bfseries We report the first direct observation of a disordered vortex matter phase existing near the edge of a bulk type-II superconductor Nb using a novel position-sensitive neutron diffraction technique.  This ``edge-contaminated'' vortex state was implicated in previous studies using transport techniques and was postulated to have played a significant role in the behavior of vortex dynamics in a wide range of type-II superconductors.  It is found that upon thermal annealing, the vortex matter in the bulk undergoes re-ordering, suggesting that the edge-contaminated bulk vortex state is metastable. The edge vortex state remains disordered after repeated thermal annealing, indicating spatial coexistence of a vortex glass with a Bragg glass.  This observation resolves many outstanding issues concerning the peak effect in type-II superconductors.   \normalfont
\end{abstract}
% PACS, the Physics and Astronomy
              % Classification Scheme.
%\keywords{Suggested keywords}%Use showkeys class option if keyword
               %display desired
\maketitle 
Vortex matter (VM) in type-II superconductors continues to be a subject of fascination. A longstanding issue is the nature of the ground state of the vortex lines in the presence of atomic impurities and other forms of quenched disorder acting as random pinning centers. %\cite{gia-bha,Rosenstein-Li}. 
Early calculations\cite{larkin,LO} and scaling arguments \cite{imry-ma} suggested the absence of long-range order even for weak random pinning. Thus an ordered VM phase was not expected in any real type-II superconductors contrary to the neutron diffraction experiments in which sharp Bragg peaks were observed in the VM phase of Nb \cite{Schelton,Christen}. A possible reconcilliation has since been proposed in the Bragg glass model \cite{Nattermann,GiLeDPRL} which predicts that the vortex lines form a topologically ordered lattice with {\it quasi-long-range-order (QLRO)} characterized by a power-law structure factor\cite{GiLeDPRB,Klein}. However, the latest theoretical treatments seem to again keep the issue in the open\cite{tis-tar,led-wie,Fisch} and further experimental progress is urgently needed. 

The existence of QLRO implies that there must be a true order-disorder phase transition in the VM.  An outstanding question is how this putative phase transition is related to the ubiquitous {\it peak effect} \cite{Autler,DeSorbo,Kes-Tsuei,ling-budnick,kwok,ishida,Shi,nishizaki,Ling,Marchevsky2001Nature,Troyanovski,Pissas,Park,lortz,Pasquini}, an intriguing phenomenon that when disorder is reduced in sample preparation, a type-II superconductor often exhibits a sharp minimum in the resistance versus temperature ($T$), or a peak in the critical current versus $T$ curves\cite{Autler}.  The peak effect has attracted attention \cite{AK,Pippard} since it was first observed in Nb\cite{Autler,DeSorbo}, but remains poorly understood. There is clear evidence for a first-order melting transition at the peak effect from magnetization\cite{nishizaki,Shi}, neutron diffraction\cite{Ling}, and heat capacity\cite{lortz} measurements. It is also known that the peak effect can disappear at high or low magnetic fields \cite{ling1,Park,Daniilidis}, and some samples display neither a peak effect nor any sign of a phase transition\cite{Forgan2002PRL,forgan2010}. This ``lack-of-universality'' appears to be related to another puzzle in the VM physics: in some samples, the zero-field-cooled (ZFC) VM state is ordered, but the field-cooled (FC) state is disordered, indicative of supercooling at a first-order transition \cite{Ling,Pasquini,xiao}; while in other samples, the reverse is true \cite{Yaron1994PRL}.  There have been strong indications from transport studies \cite{Paltiel,xiao1} that the latter effect may be caused by an ``edge-contamination'' mechanism. In some samples, due to a highly inhomogeneous surface barrier,  ``tearing''\cite{Bhattacharya} can occur as the vortex lines are driven into system from the sample edges.   This model provides an excellent explanation for transport\cite{Paltiel,xiao1,Reibelt} and magnetization experiments\cite{kupfer}. However, there has been no direct structural evidence for the existence of an edge-contaminated VM phase. Here we report the first direct evidence that in a Nb single crystal exhibiting a disordered ZFC state, the VM near the sample edge is indeed disordered.  We show that the disordered bulk ZFC VM phase is metastable and can be thermally annealed into large ordered domains.  In contrast, the edge VM state remains disordered.  

\begin{figure}
 \includegraphics[width=3.4in]{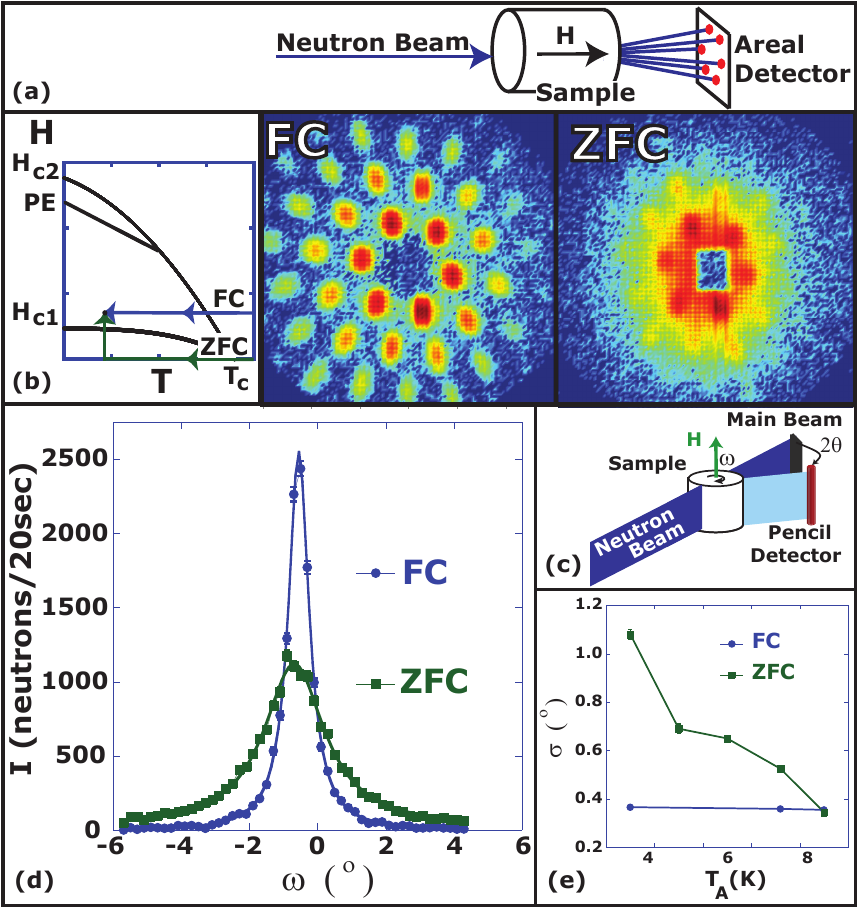}
\caption{(Color online)(a) The scattering geometry for conventional small-angle neutron scattering (SANS). (b) (Left) The magnetic field - temperature phase diagram; (Right) SANS images, summed over the rocking angle, for the FC and the ZFC states at $T=1.5$ K and $H=1400$ Oe. The neutron wavelength is \mbox{$\lambda=12$ \AA} and \mbox{$\Delta\lambda/\lambda=0.1$}. Rectangular guides are used for angular collimation with an angular spread of $0.146^\circ$. (c) The scattering geometry for AND/R. The incident neutron beam has a mean wavelength of \mbox{$\lambda=5$ \AA} and a \mbox{$\Delta\lambda/\lambda=0.01$}. Measurements of the main neutron beam reveal an angular spread of $0.02^{\circ}$ and $0.009^{\circ}$ for Figs.\ \ref{ExperimentS}, \ref{Edge}, and for Fig.\ \ref{Slicing} respectively. Note the magnetic field is applied parallel to the cylindrical axis of the sample. (d) The initial Bragg peak structure at $T=3.5$ K and $H=1400$ Oe for ZFC and FC VM. Solid lines are Lorentzian fits. (e) The Lorentzian half-width, $\sigma$, of the Bragg peaks at \mbox{$H=1400$ Oe} is plotted as a function of the annealing temperature, $T_A$. $T_A$ is the highest temperature the vortex lattice is heated to before returning to the measurement temperature, $T=3.5$ K.} \label{ExperimentS}
\end{figure}

Our experiment is made possible by a reflectometry instrument\cite{Dura} which has the novel capability of neutron diffraction topography\cite{Schlenker1975}. This technique allows positional-dependent structural analysis of a bulk vortex state in a type-II superconductor. Our sample is an as-grown Nb single crystal\cite{Goodfellow} of $99.99\%$ in purity.  It has the shape of a cylinder with the $\left\langle 111\right\rangle$ crystallographic direction oriented along its cylindrical axis. The crystal diameter is $12.1$ mm with a height of $10.1$ mm and weighs $9.69$ g. From our AC magnetic susceptibility measurements (data not shown), we sketch the magnetic phase diagram for this sample in Fig.\ \ref{ExperimentS}.  The important sample parameters are: by measurement, the zero-field superconducting transition temperature \mbox{$T_c = 9.2$ K}, and the peak effect \mbox{$H_p (4.2K)$ = 4000 Oe}; by extrapolation, the zero-temperature critical field values \mbox{$H_{c1} = 900$ Oe}, \mbox{$H_{c2} = 5800$ Oe}, and the zero-temperature peak effect \mbox{$H_p = 4800$ Oe}. These values are similar to those of a Nb crystal used in previous studies\cite{Ling, Park, Daniilidis}. However, there are two notable differences between these two samples.  First, this Nb crystal has a residual-resistivity-ratio (between 300 K and 10 K) of RRR=32, while the previous one has RRR=14\cite{Daniilidis}. Second, in the previous sample, the peak effect was pronounced and observable (using AC magnetic susceptibility\cite{ling-budnick}) down to 800 Oe\cite{Park}, while in the present sample, a remnant of the peak  effect is observable in temperature-dependent AC susceptibility measurements at fields above \mbox{3300 Oe}, and a weak peak effect is clearly observed at \mbox{4.2 K} and \mbox{4000 Oe}.  Below we use neutron diffraction techniques to unravel the vortex physics underlying the absence of the peak effect at low fields.

Our thermo-magnetic experimental procedures are defined on the phase diagram in Fig.\ \ref{ExperimentS}: Field-cooled (FC) is cooling the sample in a magnetic field of \mbox{$1400$ Oe} and zero-field-cooled (ZFC) is cooling the sample in zero magnetic field and then ramping the field to \mbox{$1400$ Oe}.  Fig.\ \ref{ExperimentS} demonstrates the drastic difference in order between the FC and ZFC VM. The standard small angle neutron scattering (SANS) geometry is shown in Fig.\ \ref{ExperimentS}(top). The magnetic field, $H$, is parallel to the neutron beam and the cylindrical axis of the sample. The SANS data in Fig.\ \ref{ExperimentS}(b, Right) is measured using the V4 instrument at the BER II reactor of the Helmholtz Zentrum Berlin. The FC SANS image (center) shows well defined Bragg peaks to several orders indicating that the FC VM is clearly ordered. For the ZFC state (right), the Bragg peaks are emerging but there is significant scattering between the expected Bragg peaks indicating a disordered vortex state.  Due to the present technical limitation on most SANS instruments, it is difficult to characterize the spatial variation of the VM structure.  Below we show that one can overcome this difficulty by using a neutron reflectometor in the diffraction mode. 

\begin{figure}
 \includegraphics[]{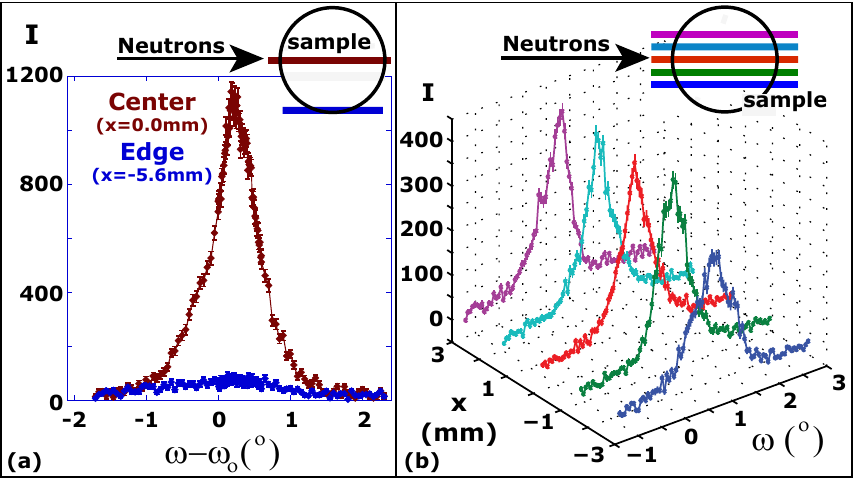}
\caption{(a)(Top) A top view of the location to the two Bragg peaks. (Main) Intensity, $I$, versus rocking angle minus the location of the Bragg condition, $\omega-\omega_{\circ}$, of the Bragg scattering from the edge of the sample and the scattering from the sample center for an AFC VM scaled at \mbox{$4.5$ K} and \mbox{$1400$ Oe}. (b)3D plot of the AFC Bragg peaks at $4.5$ K and $1400$ Oe: I versus $\omega$ versus sample position ($x$) with a top view of the respective spatial positions in the Nb crystal. The solid lines are guides for the eyes and all intensities are scaled to the same scattering area.}
\label{Slicing}
\end{figure}

At the NIST-Center for Neutron Research, we use the Advanced Neutron Diffractometer/Reflecter (AND/R)\cite{Dura} for diffraction as shown in Fig.\ \ref{ExperimentS}(c). This allows us to investigate the radial and azimuthal widths of the in-plane Bragg peak as the sample is scanned by a ribbon-shaped neutron beam. The azimuthal direction is measured by rotating the sample (and the magnetic field) and measuring the scattered neutron intensity as a function of rotation angle, $\omega$. A single detector is set at the angle $2\theta$, where $\theta$ satisfies the Bragg condition ($n\lambda=2d\sin\theta$, n=1 and $d$ is the vortex plane spacing). The magnetic field is applied parallel to the $\left\langle 111\right\rangle$ axis of the Nb sample, perpendicular to the neutron beam collimation. The AND/R instrument allows us to explore the spatial nature of the disorder in the VM. The neutron beam with a width of $0.5$ mm is much smaller than the sample diameter ($12.1$ mm). We are able to vary the section of the Nb crystal exposed to the neutron beam and measure the Bragg peak for a particular spatial location (the drawn-to-scale sketch of the topography measurement is shown in Fig.\ \ref{Slicing}). However, the increased resolution is at the expense of neutron flux, and prevents us from studying the peak effect and the order-disorder transition directly. Our scanning neutron diffraction measurements are limited to deep in the Bragg glass portion of the phase diagram. For probing different sections of the VM, we calibrate the center position ($x=0.0$ mm) via scans of a neutron absorber (Cd mask) located on the bottom of the sample. 
 
Fig.\ \ref{ExperimentS}(d) shows typical AND/R data, scattering intensity vs. rocking angle, for the same field 1400 Oe and growth histories at $T=3.5$ K, as in the SANS measurements. It can be seen from Fig.\ \ref{ExperimentS}(d) that $\sigma_{\mbox{\scriptsize{FC}}}<\sigma_{\mbox{\scriptsize{ZFC}}}$, where $\sigma$ is the Lorentzian half-width of the peak. The broadness and asymmetry in the ZFC peak suggest a vortex lattice broken into many small domains separated by low-angle grain boundaries. The sharp FC peak is not surprising since the measurement is deep inside the Bragg glass phase and no supercooling effect is expected\cite{Ling,xiao}. 

We find that the disordered ZFC state is metastable, it re-orders upon thermal annealing.  In Fig.\ 1(e), the Lorentzian half-widths of the Bragg peaks are plotted as a function of the annealing temperature, $T_A$, the highest temperature excursion the sample had undergone. It should be emphasized that all measurements in Fig.\ \ref{ExperimentS}(e) are at the same field \mbox{$1400$ Oe} and temperature \mbox{$3.5$ K}. It is clear that significant re-ordering has occurred during the annealing process for the ZFC VM.  It is interesting to note that there is a pronounced the plateau effect at $\sim T_A=6K$ which may indicate two types of re-ordering processes. We also apply the same procedure to the FC VM.  There also appears to be increased order at elevated $T_A$.  To within our experimental resolution, the ZFC and FC states approach the same final state after annealing.  
 
\begin{figure}
\includegraphics[width=3.4in]{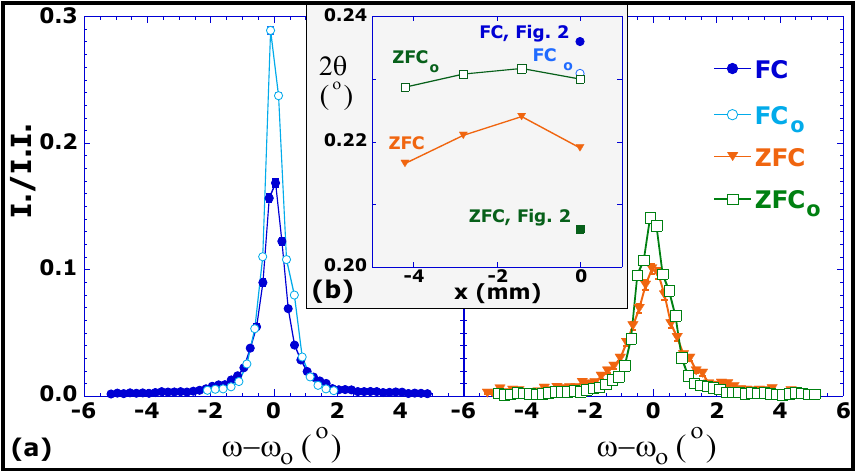}
\caption{(color online) The impact of surface oxidation on the FC and ZFC vortex matter is shown. (a) The intensity divided by the integrated intensity (I./I.I.)  is plotted versus $\omega-\omega_{\circ}$ for the FC versus FC$_o$ VM (right) and ZFC versus ZFC$_o$ VM (left). (b)The $2\theta$ value is plotted versus the sample position, $x$, for ZFC and FC VM. The measurements are at \mbox{$T=4.5$ T} and \mbox{$H=1400$ Oe}.}
\label{Edge}
\end{figure}
Next we use AND/R to characterize the spatial variation of the annealed VM phase.  Fig.\ \ref{Slicing}(a) shows the rocking curves from the center and edge of the sample for an annealed FC state.  A sharp asymmetric peak is found for the center of the sample indicating an ordered VM phase.  Upon a close examnination, one can discern fine structures indicating large ordered domains of VM separated by low angle grain boundaries.  We interpret this state as the Bragg glass phase with residual nonequilibrium effects.  The broad Bragg peak at the edge of the sample suggests a disordered VM phase even after annealing.  We suggest that this may be a vortex glass phase \cite{ffh} co-existing on the sample edge.  Fig.\ \ref{Slicing}(a) is thus the direct structural evidence for the edge contamination model. In Fig.\ \ref{Slicing}(b), we measure the variation of the Bragg peak structure through the interior of the sample.  As the neutron beam moves across the sample, there are clear changes in the structure of the Bragg peak.  This indicates a complex domain structure in the VM, consistent with previous Reverse Monte Carlo simulations on SANS data\cite{Laver2008PRL}. 

Fig.\ \ref{Slicing}(a) shows that the disordered edge VM exists even in the annealed state. This observation is significant since it implies that the structural information obtained by standard SANS measurements is not directly correlated with dynamical measurements such as magnetization and transport for samples with edge contamination (with an inhomogeneous surface barrier). The SANS signal is dominated by the central region of the sample, while the latter are determined by supercurrent or magnetic field gradient on the sample surface.  This resolves an important issue raised by a recent SANS experiment\cite{Pautrat2009PRB} in which ordered VM signals can be detected by SANS above the peak effect determined by magnetization.  Furthermore, our observations offer a resolution to the longstanding puzzle concerning the appearance or absence of the peak effect. By using AC magnetic susceptibility measurements, we find that in this sample, the peak-effect line has a significantly lower slope than that of $H_{c2}(T)$. In the previous sample, the line is nearly parallel to the $H_{c2}(T)$ line. Since the AC magnetic susceptibility measurements are dominated by the surface current in the sample, the appearance or the absence of the peak effect will depend on the residual short-range order in the edge VM. If the edge VM is strongly disordered by the surface defects, there will be no observable effect from the thermally-induced disordering due to the Bragg glass melting transition in the bulk. In systems with minimal contamination effects, the peak effect is pronounced and coincides with the Bragg glass melting transition measured with SANS\cite{Ling}. An earlier observation that the peak effect in 2H-NbSe$_{2}$ can be lowered (in field for fixed $T$ measurement) by physically cutting the sample\cite{Ling2} may also be explained by the presence of a dirty vortex state at the sample edges.

Surface oxidation has been shown to suppress the surface barrier effects in Nb \cite{sekula, DerMeyPhysicaBC}. To carry out a preliminary study of the surface oxidation effect on the VM, we subject our sample to an O$_2$ atmosphere at $600^{\circ}$ C for six minutes following the recipe by Ref.\ \cite{DerMeyPhysicaBC}. After oxidation, we repeat our neutron diffraction measurements. A subscript $o$ is added to the growth procedure to denote the oxidized sample. Fig.\ \ref{Edge}(a) is a plot of the normalized intensity vs. rocking angle for the FC and ZFC states before and after oxidation. The ZFC Bragg peak has strong variations in the tail, a broader width, and a lower peak intensity. The broad tails are attributed to a strongly disordered edge state that the neutron beam bisects due to the scattering geometry. The change in the tail behavior suggests that the order in the VM near the sample edge has increased after the oxidation of the cylindrical surface. The increase in relative intensity and decrease in width in the ZFC$_{o}$ is attributed to an improved order in the bulk vortex structure. This change might be influenced by the oxidation of the top and bottom of the sample which reduces pinning of the vortex lines there and would influence the bulk behavior. Comparing FC to FC$_{o}$, we see a higher intensity in the FC$_o$ Bragg peak with minor difference in the tail supporting our conjecture. The variation in the ZFC tails is the dynamical impact of the edge contamination.

Due to the vortex pinning in the edge and bulk, the magnetic field is expected to have a spatial gradient in the ZFC state known as the Bean critical state\cite{Bean}. This is confirmed in our radial measurements.  As plotted in Fig.\ \ref{Edge}(b) there is a difference of $0.03^{\circ}$ between the $2\theta$ of the ZFC and FC VMs which corresponds to a change in the vortex plane spacing and a field difference of \mbox{$300$ Oe}. Fig.\ \ref{Edge}(b) is a plot of $2\theta$ versus sample location, $x$, and shows that the surface oxidation has reduced the ZFC VM field profile, as expected from a reduced surface barrier. 

In summary, using a novel slicing neutron diffraction approach, we have explored the VM in a type-II superconductor with edge contamination. Our data offer the first direct evidence that the edge contamination is indeed present in systems with a disordered ZFC vortex state. Our results shed new light on the peak effect problem in type-II superconductors and may offer a new route to the growth of a true Bragg glass state.

This research was supported by the U.S. Department of Energy, Office of Basic Energy Sciences, Division of Materials Sciences and Engineering under grant DE-FG$02-07$ER$46458$. The AND/R measurements were supported by the Department of Commerce. The SANS measurements were supported by the European Commission under the $7^{th}$ Framework Programme through ``Research Infrastructures'' action of the ``Capacities'' Programme, contract number CP$-$CSA INFRA$-2008-1.1.1.$ Number $226507-$NMI$3$. H.A.H. and X.W. would like to acknowledge support from the Galkin Fund at Brown University.

\footnotesize
$^\ast$Current address: Brookhaven National Laboratories, Upton,
NY 11973, USA.

$^\dagger$Permanent address: School of Physics, Wuhan University,
Wuhan 430072, PRC.
\bibliographystyle{unsrt}

\begin{thebibliography}{50}
\scriptsize
\setlength{\itemsep}{0mm}
%\bibitem{gia-bha}
%T. Giamarchi and S. Bhattacharya, in ``High Magnetic Fields: Applications in Condensed Matter Physics and Spectroscopy", p. 314, ed. C. Berthier et al., Springer-Verlag, 2002.
%
%\bibitem{Rosenstein-Li}
%B. Rosenstein and Dingping Li, Rev. Mod. Phys. \normalfont{\bf
%82}, \normalfont 109 (2010).

\bibitem{larkin}
A.I. Larkin, Sov. Phys. JETP \normalfont{\bf 31}, 784 (1970).

\bibitem{LO}
A.I. Larkin and Yu.N. Ovchinnikov, J. Low Temp. Phys. {\bf 34}, 409 (1979). 

\bibitem{imry-ma}
Y. Imry and S. Ma,
Phys. Rev. Lett. \normalfont{\bf 35}, 1399 (1975).
%
%\bibitem{Abrikosov}
% A. A. Abrikosov, Soviet Physics JETP {\bf 5},\normalfont 1174 (1957).

\bibitem{Schelton}
J. Schelton, H. Ullmaier, and W. Schmatz, Phys. Status Solidi {\bf 48}, 619 (1971). 

\bibitem{Christen}
D.K. Christen {\it et al.}, Phys. Rev. B {\bf 15}, 4506 (1977).

\bibitem{Nattermann}
T. Nattermann, Phys. Rev. Lett. \normalfont{\bf 64}, \normalfont
2454 (1990).

\bibitem{GiLeDPRL}
T. Giamarchi and P. Le Doussal, Phys. Rev. Lett. {\bf 72}, 1530 (1994).

\bibitem{GiLeDPRB}
T. Giamarchi and P. Le Doussal, Phys. Rev. B \normalfont{\bf 52},
\normalfont 1242 (1995).

\bibitem{Klein}
T. Klein {\it et al.}, Nature {\bf 413}, 404 (2001).

\bibitem{tis-tar}
M. Tissier and G. Tarjus, Phys. Rev. Lett.
{\bf 96},  087202 (2006).

\bibitem{led-wie}
P. Le Doussal and K.J. Wiese, Phys. Rev. Lett.
\normalfont\bf{96,} \normalfont 197202 (2006). 

\bibitem{Fisch}
R. Fisch, Phys. Rev. B {\bf 79}, 214429 (2009).

\bibitem{Autler}
S.H. Autler, E.S. Rosenblum, and K. Gooen, Phys. Rev. Lett. {\bf 9}, 489 (1962).

\bibitem{DeSorbo}
W. DeSorbo, Rev. Mod. Phys. {\bf 36}, 90 (1964). 

\bibitem{Kes-Tsuei}
P.H. Kes and C.C. Tsuei, Phys. Rev. B {\bf 28}, 5126 (1983). 

\bibitem{ling-budnick}
X.S. Ling and J.I. Budnick, in Magnetic Susceptibility of Superconductors and Other Spin Systems,R. A. Hein et al. (Plenum Press, New York, 1991), p. 377. 

\bibitem{kwok}
W.K. Kwok et al., Phys. Rev. Lett. {\bf 73}, 2614 (1994).

\bibitem{ishida}
T. Ishida, K. Okuda, and H. Asaoka, Phys. Rev. B \normalfont{\bf 56}, \normalfont 5128
(1997).

\bibitem{Shi}
J. Shi {\it et al.}, Phys. Rev. B  {\bf 60}, R12593 (1999).

\bibitem{nishizaki}
T. Nishizaki {\it et al.}, Physica C  {\bf 341-348}, \normalfont  957 (2000).

\bibitem{Ling}
X.S. Ling {\it et al.}, Phys. Rev. Lett. {\bf 86}, \normalfont 712 (2001).

\bibitem{Marchevsky2001Nature}
M. Marchevsky, M. Higgins, and S. Bhattacharya, Nature {\bf 409}, \normalfont 591 (2001).

\bibitem{Troyanovski}
A. M. Troyanovski {\it et al.}, Phys. Rev. Lett. {\bf 89}, \normalfont 097002 (2002).

\bibitem{Pissas}
M. Pissas {\it et al.}, Phys. Rev. Lett. {\bf 89}, 097002 (2002).

\bibitem{Park}
S. R. Park {\it et al.}, Phys. Rev. Lett. {\bf 91}, \normalfont 167003 (2003).

\bibitem{lortz}
R. Lortz {\it et al.}, Phys. Rev. B  {\bf 74}, \normalfont  104502 (2006).

\bibitem{Pasquini}
G. Pasquini {\it et al.}, Phys. Rev. Lett. {\bf 100}, \normalfont 247003 (2008).

\bibitem{AK}
P.W. Anderson and Y.B. Kim, Rev. Mod. Phys. {\bf 36}, 39 (1964) 

\bibitem{Pippard}
A.B. Pippard, Philos. Mag. {\bf 19}, 217 (1969). 

\bibitem{ling1}
X.S. Ling, J.I. Budnick, and B.W. Veal, Physica C {\bf 282}, 2191 (1997). 

\bibitem{Daniilidis}
N.D. Daniilidis {\it et al.},  Phys. Rev. B {\bf 75}, 174519 (2007).

\bibitem{Forgan2002PRL}
E.M. Forgan {\it et al.}, Phys. Rev. Lett. {\bf 88}, \normalfont 167003 (2002).

\bibitem{forgan2010}
C.J. Bowell {\it et al.}, Phys. Rev. B {\bf 82}, \normalfont 144508 (2010).

\bibitem{xiao}
Z. L. Xiao {\it et al.}, Phys. Rev. Lett. {\bf 92}, 227004 (2004). 

\bibitem{Yaron1994PRL}
U. Yaron {\it et al.}, Phys. Rev. Lett. {\bf 73}, \normalfont 2748 (1994).

\bibitem{Paltiel}
Y. Paltiel {\it et al.}, Phys. Rev. B {\bf 58}, R14763 (1998). 

\bibitem{xiao1}
Z.L. Xiao {\it et al.}, Phys. Rev. B {\bf 65}, 094511 (2002). 

\bibitem{Bhattacharya}
S. Bhattacharya \& M.J. Higgins, Phys. Rev. B {\bf 52}, \normalfont 6467 (1995).

\bibitem{Reibelt}
M. Reibelt, A. Schilling, and N. Toyota, Phys. Rev. B {\bf 81}, 094510
(2010).

\bibitem{kupfer}
H. K\"{u}pfer {\it et al.}, Phys. Rev. B {\bf 70}, \normalfont 144509
(2004).

\bibitem{Dura}
J. A. Dura {\it et al.}, \normalfont Rev. of Sci. Instr.
\normalfont\bf{77, }\normalfont 074301 (2006).

\bibitem{Schlenker1975}
M. Schlenker, J. Baruchel, R. Perrier de la B\^{a}thie, J. Appl. Phys. {\bf 46}, 2845 (1975).

\bibitem{Goodfellow}
The crystal was commercially grown by Goodfellow Ltd. UK. 

\bibitem{Li2006PRL}
G. Li {\it et al.}, Phys. Rev. Lett. {\bf 96}, 107009 (2006).

\bibitem{ffh}
D.S. Fisher, M.P.A. Fisher, and D.A. Huse, Phys. Rev. B {\bf 43}, 130 (1991).

\bibitem{Laver2008PRL}
M. Laver {\it et al.}, Phys. Rev. Lett. {\bf 100}, 107001 (2008).

\bibitem{Pautrat2009PRB}
A. Pautrat {\it et al.}, Phys. Rev. B {\bf 
79}, 184511  (2009).

\bibitem{Ling2}
X.S. Ling {\it et al.} Phil. Mag. Lett. {\bf 79}, 399 (1999).

\bibitem{sekula}
S.T. Sekula and R.H. Kernohan, Phys. Rev. B {\bf 5}, 904 (1972).

\bibitem{DerMeyPhysicaBC}
G.P. Van Der Mey, P.H. Kes, and D. de Klerk, Physica B, {\bf 95}, 369 (1978).

\bibitem{Bean}
 C.P. Bean, Phys. Rev. Lett., {\bf 8}\normalfont\, 250 (1962). 



\end{thebibliography}

\end{document}